# Memetic evolution of art to distinct aesthetics

Francis AM Manno III

*Abstract*— Humans have been using symbolic representation (i.e. art) as a creative cultural form indisputably for at least 80,000 years (Bouzouggar, et al., 2007. PNAS Jun 12;104(24):9964-9). A description of the processes central to the evolution of art from sculpted earthen forms early in human existence to paintings in museums of the modern world is absent in scholarly literature (Gombrich, 1972; Spivey, 2005). The present manuscript offers a memetic theory of art evolution demonstrating typological transitions of art in the period post the 3-dimensional to 2-dimensional transition occurring in the early Aurignacian period (Pike, et al., 2012. Science. 2012 Jun 15;336(6087):1409-13). The process of art evolution in the 2-dimensional form central to the typological transitions post this period was propelled by material and/or structure chosen to display the artistic message. Symbolic representational 'tinkering' (Jacob, 1977) whether collectively or individually creates artistic typologies of display deemed aesthetic and non-aesthetic by the current cultural milieu. Contemporary forms of graffiti underscore the aesthetic/nonaesthetic dichotomy.

## I. Introduction

THE etiology of art on a 2-dimensional surface has its indelible origins in the caves of Altamira, Spain based on uranium dating (Pike et al, 2012; also in the caves of Chauvet (Sadier, et al., 2012) & Lascaux, France and elsewhere in Europe and Australia). As an aside, I would argue the beginning of 2-dimensional art was when our ancestral homind hunter-gathers spilled blood on their hide clothing, none of this meme, insofar as I know, survives today from the prehistoric peoples. To-date, it is believed humans have been creating art (symbolic representation) for approximately 80,000 years (Bouzouggar, et al., 2007), although some sources claim longer (Appenzeller, 1998; Bednarik, 1994; 2003). Contemporary descriptions of art lack an adequate account of the evolutionary origins of symbolic representation to distinct typological forms accounting from prehistoric times to current artistic expressions (Gombrich, 1972; Spivey, 2005). Current research has focused on ethnographic (Lewis-Williams, 2004) archeological (Bednarik, 1994; 2003), and neuroaesthetic (Zeki 2002; Ramachandran, Hirstein, 1999) mechanisms which may fundamentally underlie the evolution of art. While these analytic methods offer a constructive approach, there is still much to be ascertained as to how symbolic representation evolved.

The subject of the present manuscript approaches the evolution of 2-dimensional symbolic representation from a memetic vantage. From the first 2-dimensional symbolic representations painted in the caves of Altamira, Spain (Pike et al, 2012) and Chauvet, France (Sadier, et al., 2012) to contemporary art (Spivey, 2005), an account of the evolution of this memetic form is lacking. To describe the evolution, genetic constructs and principles were used to assemble a theory of how a particular art meme evolved to different typological forms (Figure 1). The art meme tackled here is the material used to represent the symbolic expression (i.e. what the painted message was displayed on: cave, rock, building wall, canvas, etc.). Different forms of this material/structural art meme have evolved from the Paleolithic three-dimensional sculpted forms (Bednarik, 1994; Bednarik, 2003; Bouzouggar, et al., 2007) to the advent of cave paintings (Pike, et al. 2012) to contemporary artistic expressions painted on canvas and graffiti on building walls. The evolutionary processes involved appear to be propelled by collective and/or individualistic 'tinkering' (Jacob, 1977; Figure 2, Figure 3 G, H, I). Tinkering can lead to drastically nonaesthetic forms of art as deemed by the current cultural milieu. Nonaesthetic art is described here as an art form lacking comparative value to similar art forms. The theoretical explanation ascribed herein is not the endgame for how symbolic representation evolved; moreover it's implied that this theory is one approach to deciphering current aesthetic forms seen in our everyday lives.

## II. Deciphering the evolution of 2-dimensional art

To decipher the evolution of 2-dimensional art, three objectives were posited: 1) construct the memetic (transitions through culture) evolution of a typical art form to distinct typologies, 2) demonstrate distinct typological transitional forms for the particular meme, and 3) suggest how evolution created aesthetic and nonaesthetic art by tinkering which may, with additional assistance, be deemed aesthetic (i.e. certain forms of graffiti). The following definitions are largely based on biological relationships (Figure 1; Fitch, 2000). The concept of a meme was first described by Richard Dawkins in The Selfish Gene as a "new replicator, a noun that conveys the idea of a unit of cultural transmission, or a unit of imitation" (Dawkins, 1976:192). In biology, Characters are "any genic, structural, or behavioral feature of an organism having at least two forms of the feature called character states for example red (cardinals) or blue (blue jays)" (Fitch, 2000). For cultural memetics pertaining to art, Characters – are any memetic,

Instituto de Neurobiología UNAM, Campus Juriquilla, Boulevard Juriquilla 3001, Juriquilla, Querétaro. C.P. 76230 México (email: Francis.Manno@nyu.edu): The author would like to thank Francis A. Manno Jr. and Helen D. Manno for their support, and acknowledge the Instituto de Neurobiología at the Universidad Nacional Autónoma de México and his current advisor Dr. Luis Concha; for providing the opportunity to continue his erudition.

structural or behavioral feature or attribute for a piece of art having at least two forms of the feature called character states (Fitch, 2000). Examples: structure - using stencils versus a color palette (See figure 2: A-Cockerel and B-Roster, respectively) and behavior such as brushstroke rhythmicity and fractal patterns which differentiate paintings such as Vincent van Gogh (Li, et al., 2012) and Jackson Pollock (Taylor, et al., 1999), respectively. The forms of the feature in Figure 2 would be the stencils versus choice of color utilized. For the current manuscript the type of material/structure used to paint on is of importance (Figure 3). Art in the 2-dimensional form can be typologically traced to different character states (Figure 3).

rock painting and bark canvas art (a cartographic representation of the landscape, primarily used as a map, but with considerable aesthetic) to demonstrate the divergence of memes (Figure 3 B & C). These forms bridged to painting on building walls in several forms (mural, stencil and spray graffiti) to mobile forms on vans, subways, and railroad cars, and to contemporary canvas laden art we see in museums and hang on our walls at home. Figure 3 illustrates several relationships which exist with the current forms of art and their origins. Putative relationships are illustrated with dashed lines where complete information concerning that particular meme is absent.

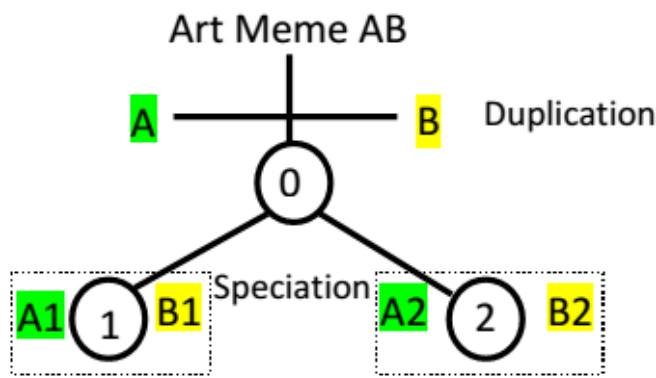

Figure 1 | Art meme cultural transmission. The 'Art Meme' can be any character such as color used on a palette, material/structure used for artistic creation (canvas, rockwall), or the behavior implemented to create the particular art, i.e. going to a mountain or cave to create a particular art. This particular 'Art Meme' has two characteristics A and B derived from a duplication event, AB leading to the different colors for the meme A color painting and B black/white painting. A speciation event 0 in the circle leads to two different characteristics 1 and 2 in circles, which for example could be the different object painted on (canvas versus rockwall) maintaining the two color characteristics. Figure 1 derived from Fitch, 2000; Jensen, 2001; Koonin, 2001.

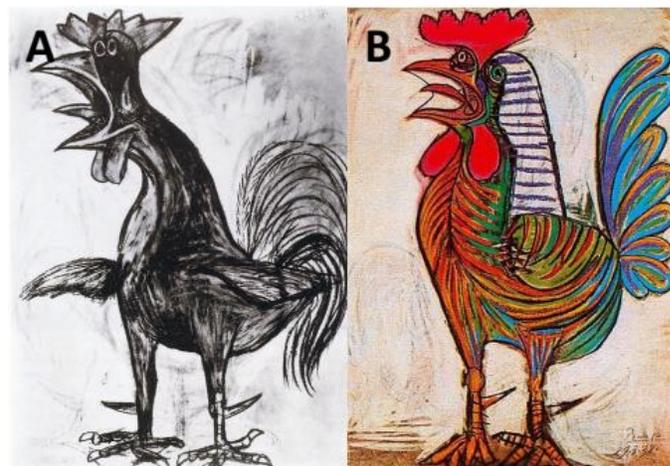

Figure 2 | From Left to right, Pablo Picasso sketches and paintings: A) Cockerel 23 March 1938, B) Rooster 27 March 1938. The Cockerel was originally juxtaposed to the hen (La mère poule) and her chicklets by Gombrich 1978;P9. The roster/cockerel illustrates Picassoian tinkering and Picasso's memetic evolution of the 'Rooster' or 'Cockerel'. The Picassoian art meme of the 'Rooster' or 'Cockerel' has become so prolific there is a children book published about Picasso's Roosters (Appel, Guglielmo, 2009). See Figure 1 for evolutionary relatedness where A and B are a duplication on the meme roster/cockerel, therefore are paralogous characters distinguished by stencil and color palette (Koonin, 2001; Jensen, 2001).

### III. AN ARGUMENT FOR ART MEMES

The present manuscript argues tinkering propelled artistic memes. An example of individualistic tinkering of an art meme by Picasso is shown in Figure 2. The roster/cockerel meme of Picasso can be characterized by two states: stencil and color palette as described by Figure 1. If we assume tinkering both individualistically and collectively over extended periods of time (i.e. Darwin, 1869; Jacob, 1977), several relationships can elucidated and ascribed to associations of symbolic representation evolving to different character states. Figure 3 demonstrates the memetic cultural transmission of an art to distinct typologies, i.e. material/structure painted on. Figure 3 is a succinct schematic including only the essential typological transitions to the memes of interest (graffiti and canvas art). The first 2-dimensional representational art appeared on the scene ≈40,000 BCE (Bednarik, 1994; Bednarik, 2003; Pike et al., 2012; Sadier, et al., 2012), therefore all 2-D art can trace its roots to this juncture. Two transitional forms were selected:

### IV. ACCOUNTING FOR GRAFFITI

Evolutionary biology allows one explanation of artistic evolution, albeit there are certainly other modes of art progression. To account for the perseverance of a particular art meme, a similar endgame must be employed: adaptation, change, survival of the fittest artistic expression (Darwin, 1869). A contemporary author, Gombrich (1972) compiled one of the modern tomes describing the progression of art in culture throughout the ages (see Chronological Charts; Gombrich, 1972:491-497). However, Gombrich's (1972) account was largely historical and descriptive, whereas here a theory was presented that may be tested for 'fit' to other art memes. Furthermore, nearly every piece of art Gombrich and others have discussed (Spivey, 2005), at least someone in the world would find beautiful (I found nothing in Gombrich's text that was not beautiful). Nevertheless, an abundance of subjectively ugly (i.e. nonaesthetic) art is left unaccountable by the current contemporary methodology based primarily on description. Where did 'ugly art' come from and how did it evolve from other art forms? The proposed theory presented here accounts for graffiti, a recurrent art meme consisting of muralistic paintings on railroad cars, under bridges, and along the sides of buildings to the scratched etchings in glass windows of bus stops and razor markings on stainless steel subcars, which society has often deemed vandalistic, nonaesthetic, and ugly. Graffiti as an artistic expression screams for explanation; nevertheless, no explanation has been given as to how this artistic meme evolved from early art forms. The manuscript presents a compelling explanation – types of aesthetic and nonaesthetic graffiti evolved from similar art forms which had their origin in the Caves of Altimura & Chauvet, and elsewhere based on memetic evolution. If you believe all graffiti is aesthetic and the explanation provided here has not accounted for 'nonaesthetic' forms, the following example should be acknowledged: If you were to wake in the morning to find your van, home, or work had a gang tag inscribed on the side of the wheel well, apartment stairs, or wall of the building respectively, as seen in the bottom left hand of Figure 3 (the author experienced the van graffiti on an autumn morning, Figure 3 I), few individuals would describe this as beautiful art at the time of the occurrence or even think twice over the cultural evolutionary aesthetic foundations of such an abstract expression. We as scientists should take issue with this fact. Graffiti as an art is beautiful, aesthetically valuable, even if possessing nonaesthetic qualities, and fore-mostly intriguing, albeit sometimes annoying, and must be understood in order to understand the evolution of hominid art in its totality.

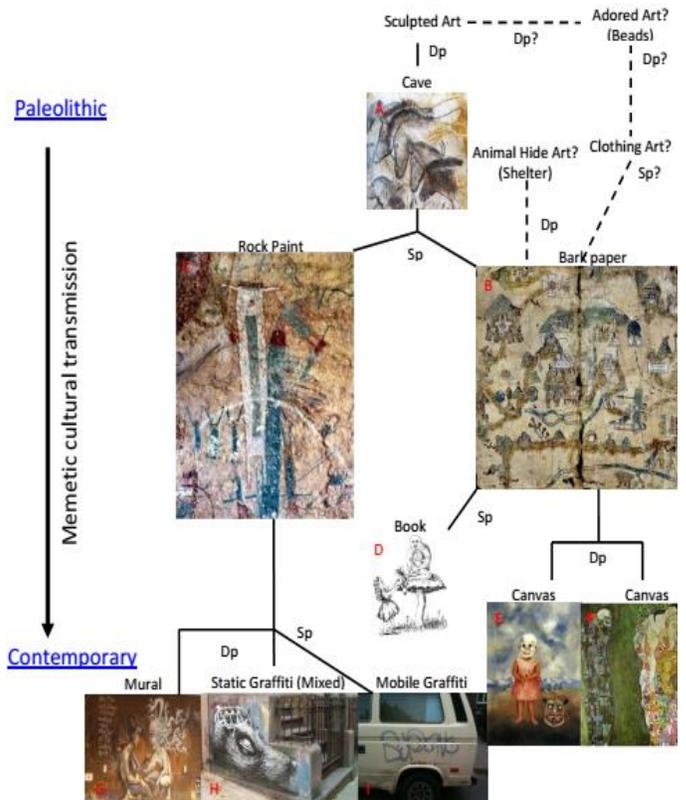

Figure 3 | Evolution of a cultural art typology: art material and/or surface - the material here is called an art meme. Solid lines show a common memetic ancestor, descending to different cultural typological outputs of that art meme. Dashed lines show putative 'hypothetical' relationships of a cultural art typology to explain possible origins. Art meme speciation events (Sp) occur at the junctions shown as an upside down Y. Art meme-duplication events (Dp), occur at a horizontal bar. Memes whose common ancestor resides at a Y junction (speciation) are orthologous. Memes whose common ancestor resides at a horizontal bar (duplications) are paralogous. Letters from A to I on pieces of art described from top to bottom. The following details the type of art, artist name, art piece name (date), and place the piece is located, separated by commas. A - Paleolithic cave art, early Human, Panel of Horses (≈40,800 BCE), Chauvet Cave, Cévennes & Rhone valleys at Vallon-Pont-d'Arc, Ardèche, France. B - Bark canvas, Mapa de Cuauhtinchan No. 2 (≈1,500 CE), Ciudad de México, Republic de México. C – Rock painting, Native Americans, therianthrope figure The White Shaman (Precise date unknown: ≈3,000 BCE), Lower Pecos Region, Seminole Canyon State Park and Historic Site, Comstock, Texas USA (National Park Service: NPS). D – Book sketching, Carroll L. (Dodgson CL), Alice & The Hookah Smoking Caterpillar (1886), Book Alice's Adventures in Wonderland Page 50. E - Canvas oil painting, Frida Khalo, Niña con Mascara de Meurte (1938), Coyoacán, Distrito Federal, Estado de México, Republic de México. F - Canvas oil painting, Gustav Klimt, Death and Life (1916), Vienna, Austria. G - Mural, Herakut, Jay (2012), 1135 Dundas East, Toronto, Canada. H – Graffiti, Le Rat (4.5.2010), Lower East Side, Manhattan, New York City, New York, USA. I - Graffiti, Gang 'Supone' (assumed writing) (2.5.2011), Side of 1982 VW Westfalia, Brooklyn, New York City, New York, USA. Considerable input from the Fitch diagram (Fitch, 2000) and commentary were utilized (Jensen, 2001; Koonin, 2001).